\documentclass[conference]{IEEEtran}

\usepackage{graphicx}
\usepackage{amsmath}
\usepackage{cite}

\begin{document}

\graphicspath{{images/}}

\title{Coupling of Wideband Impulses Generated by Granular Chains into Liquids}

\author{
\IEEEauthorblockN{Sevan Harput, and Steven Freear}
	 \IEEEauthorblockA{ \small Ultrasonics and Instrumentation Group, School of Electronic and Electrical Engineering, University of Leeds, Leeds, LS2 9JT, UK \\
	  	E-mail: S.Harput@leeds.ac.uk \\ 
	  	{ } }
	
\IEEEauthorblockN{Pierre Gelat, and Nader Saffari}
	 \IEEEauthorblockA{ \small Department of Mechanical Engineering, University College London, Torrington Place, London WC1E 7JE, UK  \\
	  	{ } }
	
\IEEEauthorblockN{Jia Yang, Omololu Akanji, Peter J. Thomas, and David A. Hutchins}
	 \IEEEauthorblockA{ \small School of Engineering, University of Warwick, Coventry CV4 7AL, UK \\ }

}

\maketitle

\begin{abstract}

An ultrasonic transducer technology to generate wideband impulses using a one-dimensional chain of spheres was previously presented. The Hertzian contact between the spheres causes the nonlinearity of the system to increase, which transforms high amplitude narrowband sinusoidal input into a train of wideband impulses. Generation of short duration ultrasonic pulses is desirable both in diagnostic and therapeutic ultrasound. Nevertheless, the biggest challenge in terms of adaptation to biomedical ultrasound is the coupling of the ultrasonic energy into biological tissue. 

An analytical model was created to address the coupling issue. Effect of the matching layer was modelled as a flexible thin plate clamped from the edges. Model was verified against hydrophone measurements. Different coupling materials, such as glass, aluminium, acrylic, silicon rubber, and vitreous carbon, was analysed with this model. Results showed that soft matching layers such as acrylic and rubber inhibit the generation of higher order harmonics. Between the hard matching materials, vitreous carbon achieved the best results due to its acoustic impedance.

\end{abstract}

\IEEEpeerreviewmaketitle

\section{Introduction}

Contact dynamics between two spherical elastic objects can be explained by Hertz's contact law with an exponential force-deformation relation. Unlike a spring mass system, where $F$ is proportional to deformation or displacement $\delta$, in a system with Hertzian contact the relation between the force $F$ exerted on spheres and the deformation $\delta$ is nonlinear; $F \propto \delta^{1.5}$. An ultrasonic transducer technology is developed to utilize this nonlinear behaviour of one-dimensional chain of spheres~\cite{Hutchins2015,Yang2016,Hutchins2016,Akanji2016}. The aim of this technology is to transform high amplitude narrowband sinusoidal input into a train of wideband impulses. The generation of short duration ultrasonic pulses is desirable both in diagnostic and therapeutic ultrasound~\cite{Bouakaz2003,Harput2014,Lin2014}.

It has already been demonstrated by Hutchins \textit{et al.} and Yang \textit{et al.} that it is possible to generate wideband impulses by coupling energy to harmonics from a fundamental ultrasonic frequency of 73 kHz~\cite{Yang2014,Hutchins2014,Hutchins2015a}. Donahue \textit{et al.} and Harput \textit{et al.} presented generation of ultrasound waves in water by using a granular chain with a matching layer~\cite{Harput2015a,Donahue2014}. The aim of this study is to incorporate the effect of a matching layer into an existing analytical model and an experimental setup to identify suitable materials and to obtain the optimum thickness for biomedical applications.

Models were developed to simulate wave propagation in infinite chains~\cite{Nesterenko1983,Coste1997}; however the problem becomes more complicated once the chain is of finite length and coupled into a finite material. Analytical models to simulate wave propagation through a finite-length chain already exist~\cite{Lydon2013,Hutchins2015}, but coupling of ultrasonic energy into biological tissue still remains as a big challenge. In this study, a matching layer was modelled as a flexible thin circular disc clamped from the edges. Different coupling materials, such as glass, aluminium, acrylic, silicon rubber, and vitreous carbon, was analysed with this model. Results achieved from the new model was verified against hydrophone measurements.

\begin{figure}[!t]
	\centering
	\includegraphics[width = 78mm]{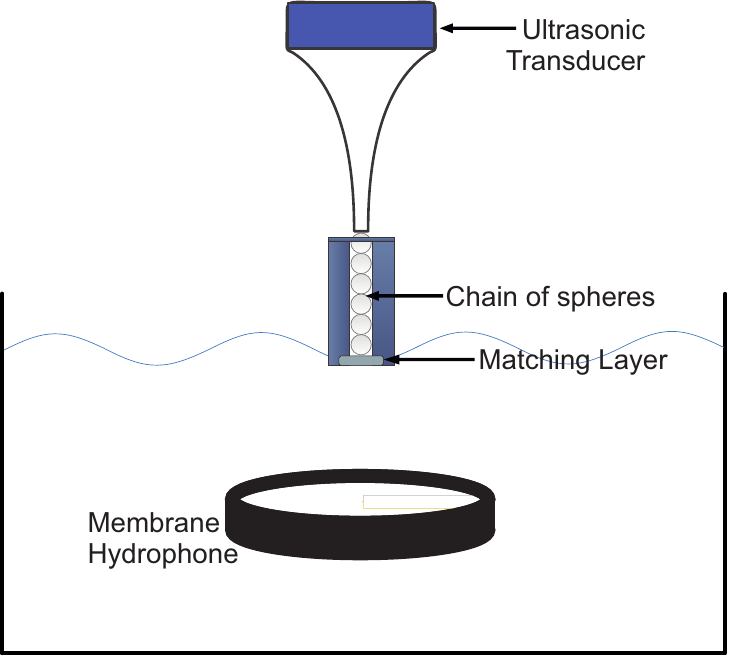}
	\caption{Chain of spheres is excited with an ultrasonic transducer attached to an amplifying horn. A matching layer is placed on the other side of the chain for coupling in water. Acoustic pressure generated by the chain is measured with a membrane hydrophone.}
	\label{fig:SoundBullets_exp_setup}
\end{figure}

\section{Materials and Methods}

\subsection{Model}
An analytical model created by Hutchins \textit{et al.} was used to simulate the wave propagation through the chain. The existing model was modified as below to implement the effect of a vibrating thin plate instead of an infinite wall. The equation from \cite{Hutchins2015} for the last sphere was modified as;
\begin{equation}
\begin{aligned}
 m \ddot{x}{_n} & = \frac{\sqrt{2R}}{3} \theta_s  (\delta_{s,s} + x_{n-1} - x_{n}) ^{\frac{3}{2}} \\
&   -  \frac{4\sqrt{R}}{3} \theta_{m,s} (\delta_{m,s} + x_{n} - x_{m})^{\frac{3}{2}} \\
&         + \lambda (\dot{x}{_{n-1}} - \dot{x}{_{n}}) H(\delta_{s,s} + x_{n-1} - x_{n}) \\
&		  - \lambda (\dot{x}{_{n}} - \dot{x}{_{m}}) H(\delta_{m,s} + x_{n} - x_{m}),
 \end{aligned}
\end{equation}
where $m$ is the mass of a single sphere, $x_n$ is the displacement of the last sphere for a $n$-sphere chain, $x_m$ is the displacement of the matching layer, and $\lambda$ is the damping coefficient. Material properties are described with $\theta_s$ and $\theta_m$ as;
\begin{equation}
\theta_s = \frac{E_s}{1 - {\nu_s}^2}, \> \> \> \> \>
\theta_m = \frac{E_m}{1 - {\nu_m}^2}
\end{equation}
\begin{equation}
\frac{1}{\theta_{m,s}} = \frac{1}{\theta_m} +  \frac{1}{\theta_s} =  \frac{E_s(1 - {\nu_m}^2) + E_m(1 - {\nu_s}^2)}{E_m E_s},
\end{equation}
where $E_s$ is the Young's modulus, and $\nu_s$ is the Poisson's ratio of the aluminium spheres. The same parameters for the matching layer are represented with $E_m$ and $\nu_m$. Overlap between two spheres, $\delta_{s,s}$, and overlap between the last sphere and the matching layer, $\delta_{m,s}$, are calculated as;
\begin{equation}
\delta_{s,s} = \left( 	\frac{3F_0}{ \sqrt{2R} \theta_s}	\right)^{\frac{2}{3}}, \> \> \> \> \>
\delta_{m,s} = \left( 	\frac{3F_0}{4 \sqrt{R} \theta_{m,s}}	\right)^{\frac{2}{3}} .
\end{equation}

The matching layer was modelled as a flexible thin plate clamped from the edges and the equation of motion is~\cite{Harput2017};
\begin{equation}
\begin{aligned}
 \frac{m_m}{4} \ddot{x}{_m} & = \frac{4\sqrt{R}}{3} \theta_{m,s} (\delta_{m,s} + x_{n} - x_{m})^{\frac{3}{2}}   \\
&		  + \lambda (\dot{x}{_{n}} - \dot{x}{_{m}}) H(\delta_{m,s} + x_{n} - x_{m}) \\
& 		  - K_m x_m ,
\end{aligned}
\end{equation}
where $m_m$ is the mass of the matching layer, and $K_m$ is the spring coefficient.
 
The mechanism of wave coupling into water was modelled as a superposition of wave propagated through the matching layer and vibrations of a thin plate. The energy of the wave transmitted from the aluminium sphere into water was decreased due to the acoustic impedance mismatch as;
\begin{equation}
T = 1 - \left( 	\frac{Z_2 - Z_1}{ Z_2 + Z_1}	\right)^2
\end{equation}
\begin{equation}
\dot{x}{_{s}} =  \dot{x}{_{n}} \sqrt{T_{m} T_{w}} H(\delta_{m,s} + x_{n} - x_{m}) + \dot{x}{_{m}}
\label{eq:vel_superpose}
\end{equation}
where $\dot{x}{_s}$ is the superposition of sphere and matching layer velocities. The first term in Eq.~\eqref{eq:vel_superpose} is the velocity of the last sphere coupled to the matching layer and then water, while the last sphere is in contact with the matching layer. The transmitted energy is scaled by $T_{m}$ into the matching layer, and scaled by $T_{w}$ into the water. The second term represents the velocity of the matching layer.

\begin{table}[!b]
\renewcommand{\arraystretch}{1.2}
	\caption{Parameters of Matching Materials} 
	\label{table:Materials}
	\centering 
		\begin{tabular}{l c c}
		\hline

		          				   	&  Young's  		& Poisson's   	  \\
		          				  	&  Modulus  		& Ratio  	  \\
		\hline
		Aluminium				  	& 69x$10^9$ Pa    			& 0.33 	   \\
		Pyrex glass 		  		& 67x$10^9$ Pa    			& 0.20 	   \\
		Vitreous carbon			  	& 35x$10^9$ Pa     			& 0.15  	   \\
		Acrylic				  		& 2.5x$10^9$ Pa     		& 0.35  	   \\
		Silicone RTV Rubber		   	& 0.1x$10^9$ Pa     		& 0.49  	   \\
		
		\hline
		
		\end{tabular}
\end{table}

\subsection{Experimental Setup}
The system as shown in Fig.~\ref{fig:SoundBullets_exp_setup} was partially submerged and the output was measured in degassed and deionized water using a PVDF membrane hydrophone (Precision Acoustics Ltd., Dorchester, UK). The chain of spheres and the ultrasonic horn was fixed on an assembly with static compression force of approximately 0.1 N. Ultrasonic horn was made of steel with a density of 7833 kg/m$^3$, Young's modulus 201 GPa, and Poisson's ratio 0.3. The chain consisted of 6 aluminium spheres with a diameter of 1 mm, density of 2700 kg/m$^3$, Young's modulus 69 GPa, and Poisson's ratio 0.33, was placed into a plastic holder. The end of the chain was terminated with different matching layers as listed in Table~\ref{table:Materials}.

The chain of spheres was excited with a 25 cycle sinusoidal tone burst generated by an ultrasonic horn with a fundamental frequency of 73 kHz. The excitation signal was tapered and amplified with a Class A linear power amplifier (E\&I Ltd., Rochester, NY) to reduce the level of harmonics introduced into the granular chain. This waveform generated a maximum output displacement of $\pm3$ $\mu$m, which was measured with a Laser Vibrometer (Polytec, Germany) and used as an input into the analytical model described above.

\section{Results and Discussion}

\begin{figure}[!t]
	\centering
	\includegraphics[viewport = 45 35 620 570, width = 82mm, clip]{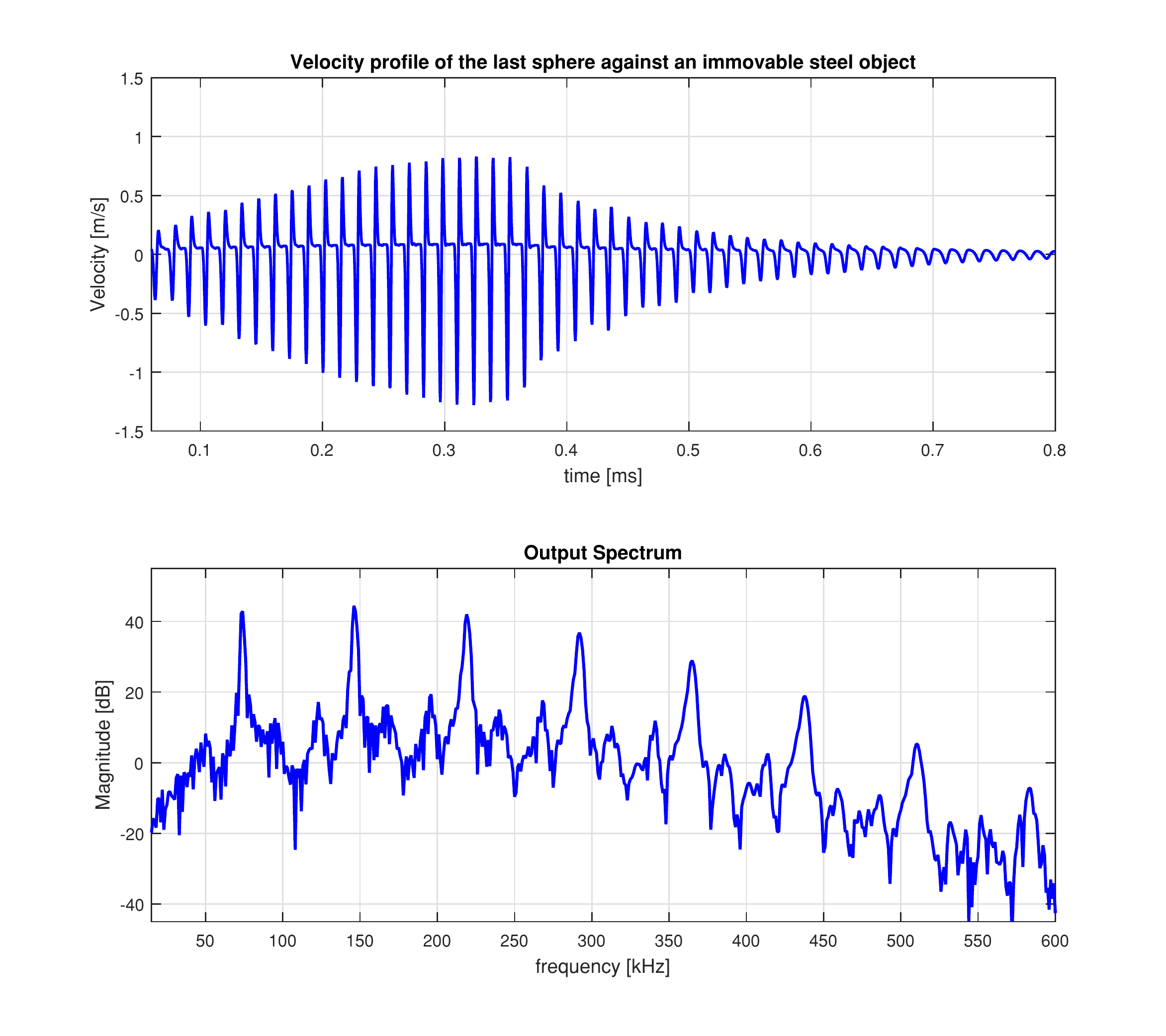}
	\caption{Results achieved with the analytical model for a 6-sphere chain against an immovable steel wall.}
	\label{fig:Simulations_Aluminium_Nocoupling}
\end{figure}

Fig. \ref{fig:Simulations_Aluminium_Nocoupling} shows the velocity of the last sphere for a 6-sphere granular chain terminated with an infinite steel wall. Chain of spheres between two immovable boundaries can generate periodic solitary waves with clear distinctions between high energy and low energy zones, which is visible in Fig. \ref{fig:Simulations_Aluminium_Nocoupling} as short duration impulses. When a non-stationary boundary is introduced, the dynamics of the chain of spheres change dramatically. The new proposed model in this study was used to simulate the effect of a thin moving boundary. Different matching materials, such as aluminium, glass, acrylic, silicon rubber, and vitreous carbon, was analysed with this model with given parameters in Table~\ref{table:Materials}. The damping coefficient was $\lambda=0.32$ as explained in~\cite{Yang2016}, the spring constant was between $K_m=2.2\times10^3-620\times10^3$ for the given materials with a pre-compression $F_0=0.1$ N.

\begin{figure}[!t]
	\centering
	\includegraphics[viewport = 45 35 620 390, width = 86mm, clip]{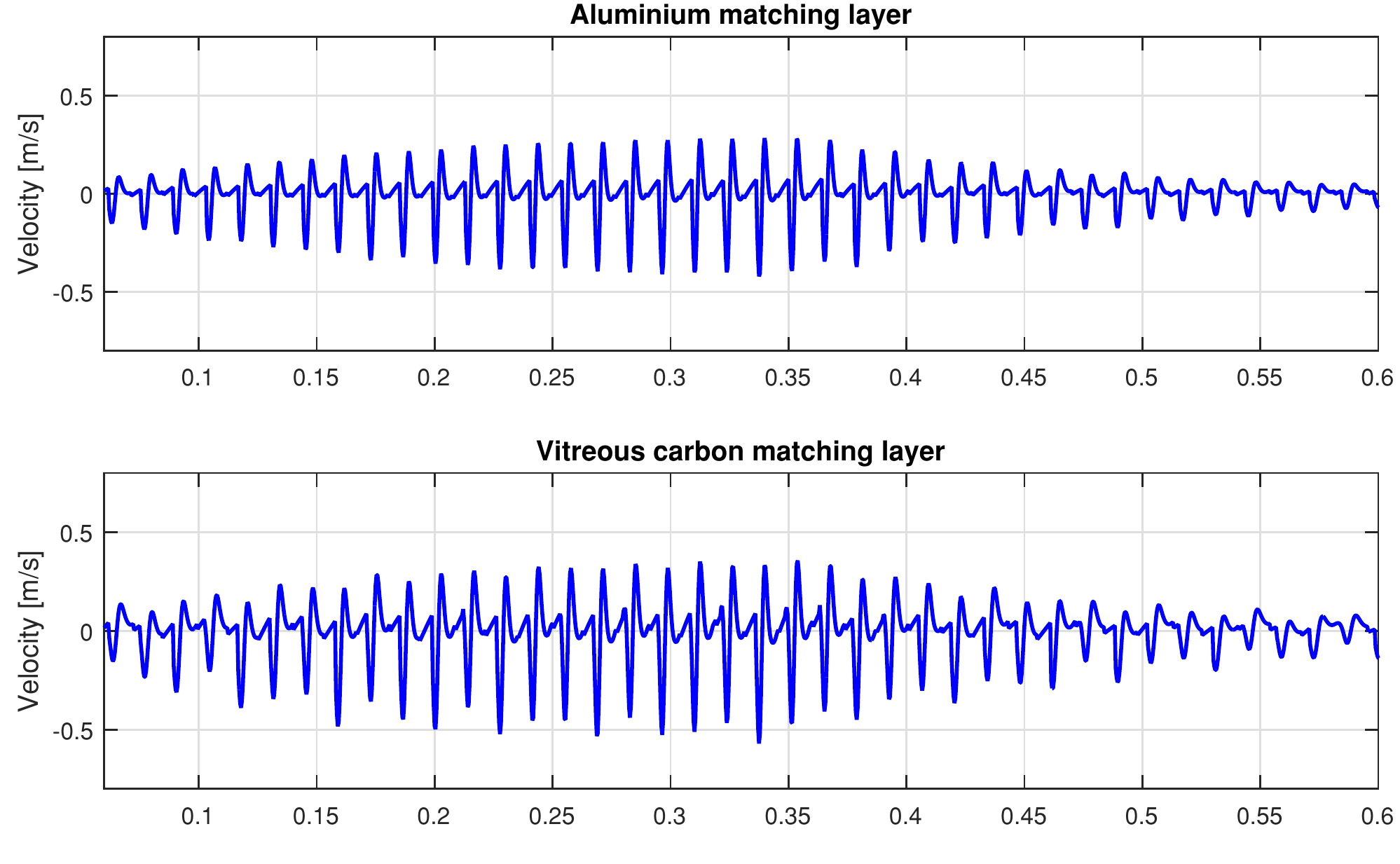}
	\includegraphics[viewport = 45 35 620 570, width = 86mm, clip]{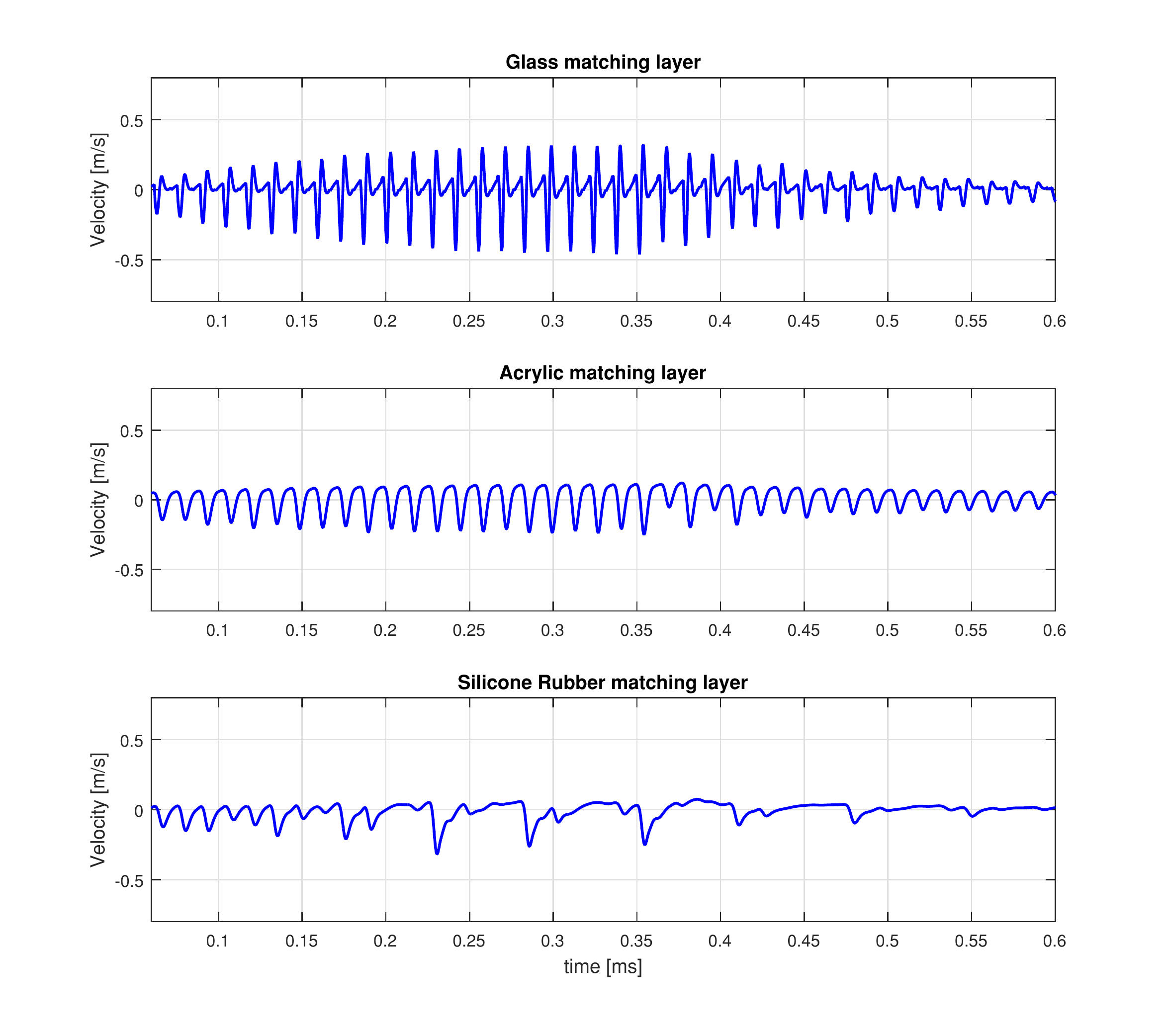}
	\caption{Results achieved with the analytical model for a 6-sphere chain coupled to aluminium, vitreous carbon, pyrex glass, acrylic, silicone rubber.}
	\label{fig:Simulations_Aluminium_ALLmatching}
\end{figure}

Fig.~\ref{fig:Simulations_Aluminium_ALLmatching} shows the estimated velocity profiles for different matching materials modelled as a circular plate with a radius of 5~mm and a thickness of 0.5~mm. Model showed that soft matching layers such as acrylic and rubber inhibit the generation of higher order harmonics. Between the hard matching materials, vitreous carbon achieved the best results due to its acoustic impedance, which is closer to the optimum impedance matching between aluminium and water. Increasing the thickness of the matching material increases the output velocity profile of the last sphere, since the system approaches to an immovable boundary with increasing thickness. However due to the existence of higher harmonics and attenuation in material, thinner matching layer is preferable.

\begin{figure}[!t]
	\centering
	\includegraphics[viewport = 45 35 620 570, width = 82mm, clip]{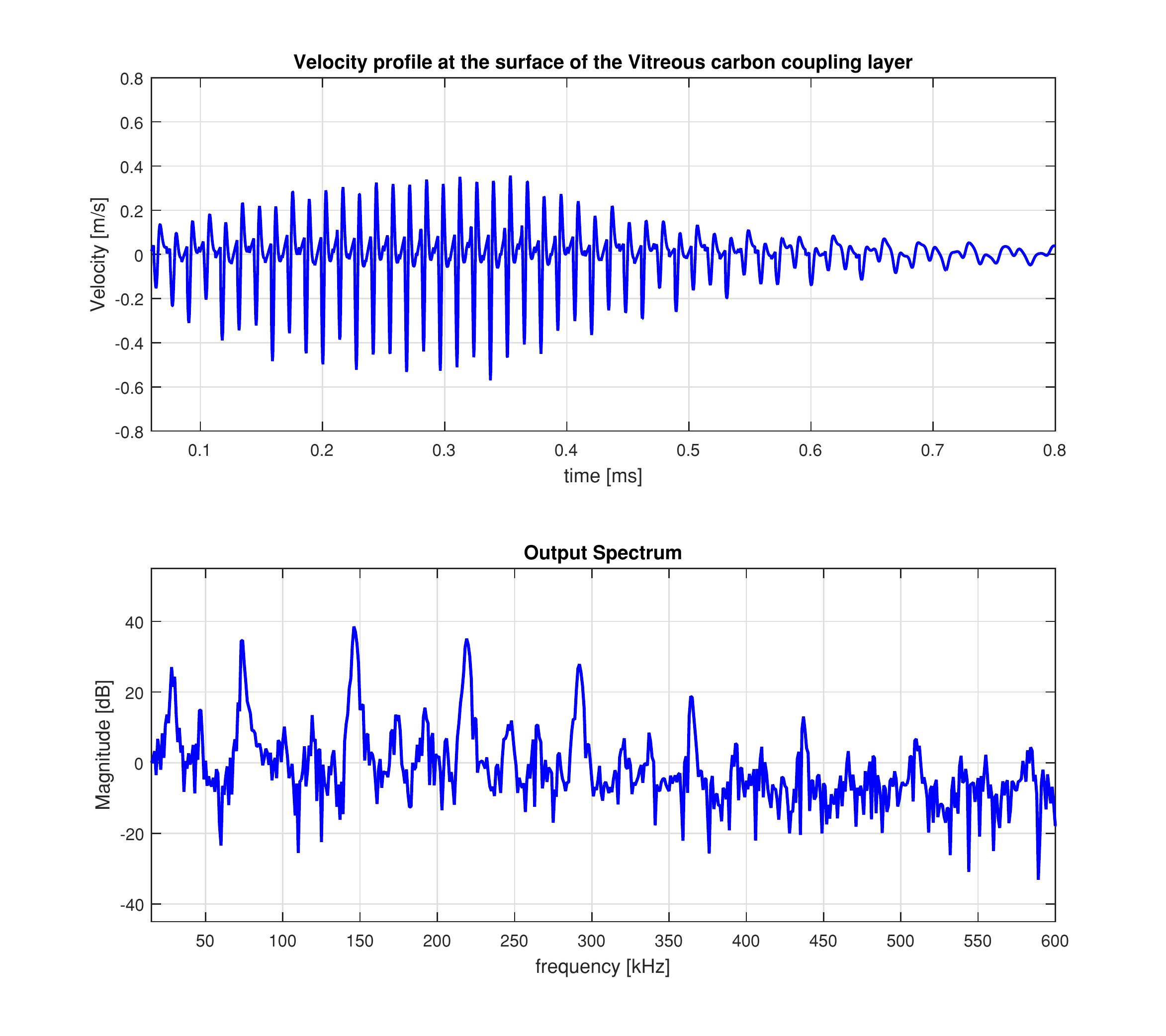}
	\caption{Results achieved with the analytical model for a 6-sphere chain coupled to a 0.5~mm thick vitreous carbon plate. (Top) Velocity profile in time domain. (Bottom) Frequency spectrum of the velocity profile. }
	\label{fig:Simulations_Aluminium_Sigradur}
\end{figure}

Fig.~\ref{fig:Simulations_Aluminium_Sigradur} shows the estimated velocity profile at the outward surface of a 0.5~mm thick vitreous carbon matching layer. Fig.~\ref{fig:Simulations_Aluminium_Sigradur}(bottom) shows the output spectrum with higher order harmonics that extended to frequencies above 400 kHz.

\begin{figure}[!t]
	\centering
	\includegraphics[viewport = 45 35 620 570, width = 82mm, clip]{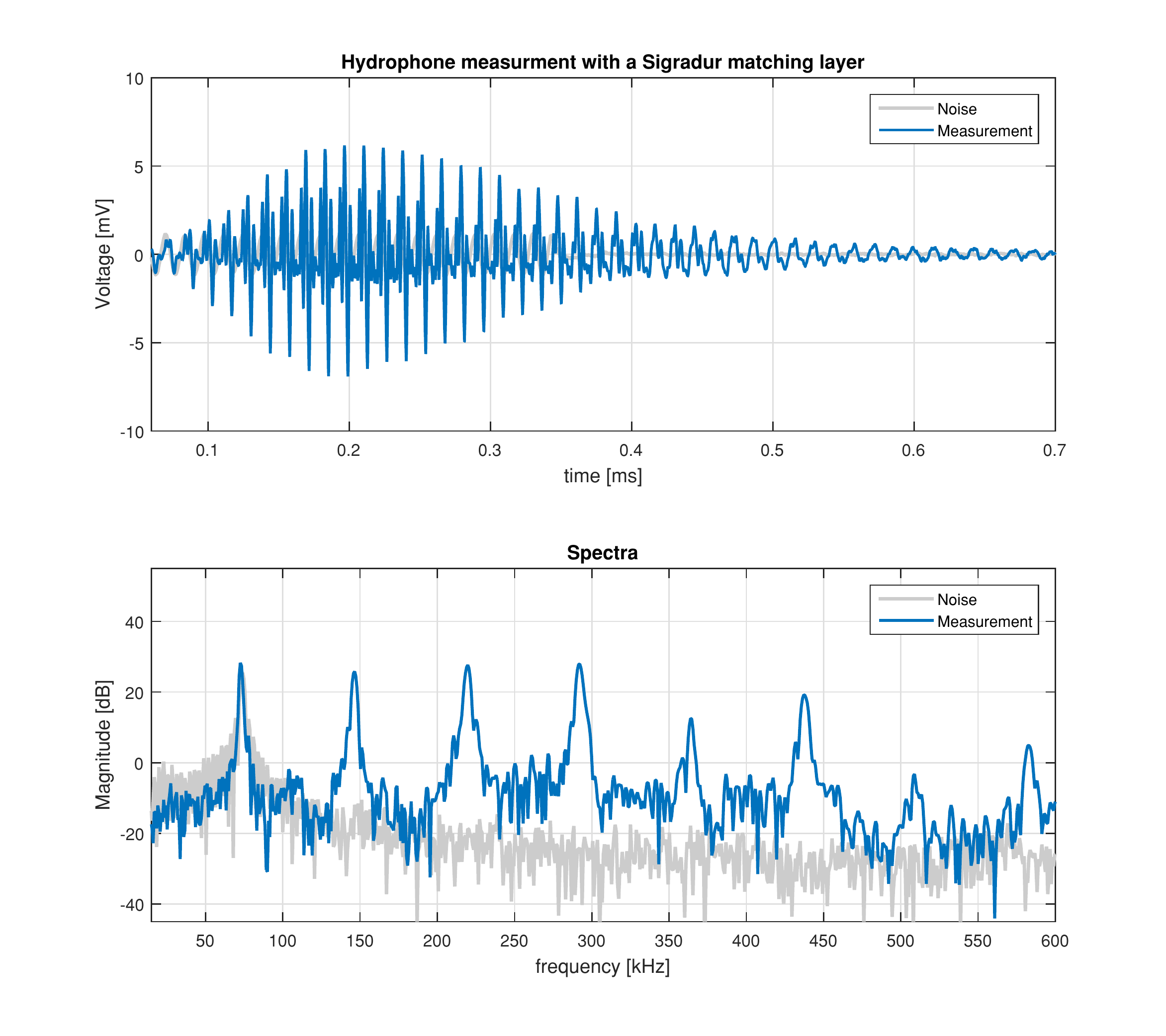}
	\caption{Measurements performed with a membrane hydrophone for a 6-sphere chain coupled to a 0.5~mm thick vitreous carbon plate. (Top) Acoustic pressure measured at 10~mm depth. (Bottom) Frequency spectrum of the acoustic pressure. Noise level is plotted for comparison.}
	\label{fig:Measurement_Aluminium_10mm}
\end{figure}

Fig.~\ref{fig:Measurement_Aluminium_10mm}(top) shows the hydrophone measurements acquired with a 0.5 mm thick vitreous carbon disc at 10 mm depth. Between 0.5, 1.0, 2.0 and 2.5 mm thick vitreous carbon discs, the thinnest option resulted in the highest output pressure that can be regarded as the most efficient coupling. It was not possible to try a thinner matching layer due to unavailability.

The predictions from the analytical model were consistent with the results achieved by hydrophone measurements for the system under static loading of 0.1 N for the vitreous carbon matching layer. The impulse shape and the harmonic content matches with the new model as shown in Fig.~\ref{fig:Simulations_Aluminium_Sigradur} and Fig.~\ref{fig:Measurement_Aluminium_10mm}. However, the envelope of the waveform is different in both cases. The reason for this is the displacement at the horn tip was effected due to the static loading.


\section{Conclusions}      

This study achieved the aim of generating wideband impulses close to biomedical ultrasound frequency range. Between the chosen matching layer materials, vitreous carbon allowed both the generation and coupling of higher order harmonics with a -20 dB bandwidth over 350 kHz and shows the feasibility of this new technology for biomedical applications. Generation and the use of these sub- and super-harmonics in ultrasound as a diagnostic and therapeutic tool are favourable~\cite{Canney2008,Harput2013,Harput2013a}. The main advantage of this new technology for therapeutic applications, such as HIFU, is the ability to generate wideband impulses that minimizes the focal spot size. The output pressure on the other hand is a big disadvantage; however it can be improved by fabricating an array of granular chains.

Thickness of the matching layer is a trade-off between efficient coupling of high frequency components (\textit{i.e.} a thin matching layer), and effective solitary wave generation ( \textit{i.e.} a thick  immovable matching layer). The current model can give a basic insight of how to choose a matching material, but the current model is incomplete. Model must be improved by adding the effect of friction and damping to the plate motion and attenuation~\cite{Harput2017}. Also for a more complicated setup with multiple matching layers, finite element analysis will be used as a supplementary tool to understand the contact mechanics in a granular chain~\cite{Gelat2016}.

\section*{Acknowledgment}     
 This work was supported by EPSRC grant (UK) EP/K029835/1.

\vspace*{1mm}
\bibliography{SoundBullets,Ultrasound,BuBBle}       
\bibliographystyle{IEEEtran}


\end{document}